\documentstyle[subfigure,preprint,aps,epsfig]{revtex}
\newcommand{\be}{\begin{equation}}
\newcommand{\ee}{\end{equation}}
\begin{document}
\draft
\preprint{\begin{tabular}{l}
\hbox to\hsize{  \hfill KAIST--TH 00/1}\\[-3mm]
\hbox to\hsize{  \hfill DESY 00--030}\\[5mm] \end{tabular} }
\title{Phenomenology of the radion in the Randall-Sundrum scenario 
\\ at colliders } 
\author{
Saebyok Bae,
~~P. Ko,
~~Hong Seok Lee
}
\address{
Department of Physics, KAIST,  
Taejon 305-701, Korea \\}
\author{Jungil Lee
}
\address{
Luruper Chaussee 149 II. Institut f\"ur Theoretische Physik  \\
   Universit\"at Hamburg, 22761 Hamburg, Germany 
}
\maketitle

\begin{abstract}
Phenomenology of a radion ($\phi$) that stabilizes the modulus in the 
Randall-Sundrum scenario is considered. 
The radion couples to the trace of energy momentum tensor of the standard 
model (SM) with a strength suppressed only by a new 
scale ($\Lambda_{\phi}$) of an order of the electroweak scale. 
In particular, the effective coupling of a radion to two gluons is 
enhanced due to the 
trace anomaly of QCD. Therefore,  its production cross section  at 
hadron colliders could be enhanced, and the dominant decay mode of a 
relatively light radion is  $\phi \rightarrow g g$, unlike the SM 
Higgs boson case. We also present constraints on the mass $m_{\phi}$ and the 
new scale $\Lambda_{\phi}$ from the Higgs search limit at LEP and perturbative 
unitarity bound. 
\end{abstract}

\narrowtext

It is one of the problems of the standard model (SM) to stabilize the 
electroweak scale relative to the Planck scale under quantum corrections, 
which is known as the gauge hierarchy problem. 
Traditionally, there have been basically two avenues to solve this problem : 
(i) electroweak gauge symmetry is spontaneously broken by some new strong 
interactions (technicolor or its relatives) or (ii) there is a supersymmetry  
(SUSY) which is spontaneously broken in a hidden sector, and superpartners 
of SM particles have masses around the electroweak scale $O(100-1000)$ GeV. 
However, new mechanisms based on the developments in superstring and M 
theories including D-branes have been suggested by Randall and Sundrum 
\cite{randsun}. If our world is confined to a three dimensional brane
and the warp factor in the Randall and Sundrum (RS) theory is much smaller
than 1, then loop corrections can not destroy the mass hierarchy 
derived from  the relation $v=e^{-k r_c \pi} v_0$,  where $v_0$ is the VEV 
of Higgs field  ($\sim$ $O(M_P)$) in the 5 dimensional RS theory, 
$e^{-k r_c \pi}$ is the warp factor, and $v$ is the VEV of Higgs field 
($\sim$ 246 GeV) in the 4 dimensional effective theory of the RS theory by 
a kind of dimensional reduction.  Especially the extra-dimensional subspace 
needs not be a circle $S^1$ like the Kaluza-Klein theory \cite{randsun}, 
and in that case, it is crucial to have a mechanism to stabilize the modulus.  
One such a mechanism was recently proposed by Goldberger and Wise (GW) 
\cite{wise}\cite{wise2}, and also by Cs${\acute{\rm a}}$ki et al.\cite{csaki}. 
In such a case, the modulus (or the radion $\phi$ from now on) is likely to 
be lighter than the lowest Kaluza-Klein excitations of bulk fields. Also its 
couplings to the SM fields are completely determined by general 
covariance in the four-dimensional spacetime, as shown in Eq. (1) below. 
If this scenario is realized in nature, this radion could be the first 
signature of this scenario, and it would be important to determine its 
phenomenology at the current/future colliders, which is the purpose of this 
work. Some related issues were addressed in Ref.~\cite{mahanta}.

In the following, we first recapitulate the interaction Lagrangian for 
a single radion and the SM fields, and calculate the decay rates and the 
branching ratios of the radion into SM particles.  Then the perturbative 
unitarity bounds on the radion mass $m_{\phi}$ and $\Lambda_{\phi}$ are 
considered.  Current bounds on the SM Higgs search can be easily translated 
into the corresponding bounds on the radion, which we will show in brief.  
Then the radion production cross sections at next linear colliders (NLC's) 
and hadron colliders such as the Tevatron and LHC are calculated. 
Then our results will be summarized at the end.

The interaction of the radion with the SM fields at an electroweak scale is 
dictated by the 4-dimensional general covariance, and  is described by the 
following effective Lagrangian \cite{csaki} \cite{wise2} :
\begin{equation}
{\cal L}_{\rm int} =  {\phi \over \Lambda_{\phi}}~{T_\mu}^{\mu} ({\rm SM})
+ ...,
\end{equation} 
where $\Lambda_{\phi} = \langle \phi \rangle \sim O(v)$. The radion becomes 
massive after the modulus stabilization, and its mass $m_{\phi}$ is a free 
parameter of electroweak scale\cite{csaki}. Therefore, two parameters 
$\Lambda_{\phi}$ and $m_{\phi}$ are required in order to discuss productions 
and decays of the radion at various settings. The couplings of the radion 
with the SM fields look like those of the SM Higgs, except for 
$v \rightarrow \Lambda_{\phi}$. 
However, there is one important thing to be noticed : the quantum corrections
to the trace of the energy-momentum tensor lead to trace anomaly, leading to
the additional effective radion couplings to gluons or photons in addition 
to the usual loop contributions. This trace anomaly contributions will lead
to distinct signatures of the radion compared to the SM Higgs boson.

The trace of energy-momentum tensor of the SM fields at tree level is
easily derived :
\begin{eqnarray}
{T_\mu}^{\mu} ({\rm SM})^{\rm tree} & = & 
\left[ \, \sum_{f} m_f \bar{f} f - 2 m_W^2 W^{+}_{\mu} W^{- \mu} - 
m_Z^2 Z_{\mu} Z^{\mu}   +  
\left( 2 m_h^2 h^2 - \partial_{\mu} h \partial^{\mu} h \right) + ... \, 
\right],
\end{eqnarray}
where we showed terms with only two SM fields, since we will discuss two 
body decay rates of the radion into the SM particles, 
except the gauge bosons of which
virtual states are also considered. 
The couplings between the radion
$\phi$ and fermion pair or weak gauge boson pair are simply related with the
SM Higgs couplings with these particles through simple rescaling : 
$g_{\phi-f-\bar{f}} = g_{h-f-\bar{f}}^{\rm SM}~{v/\Lambda_{\phi}}$, and so on.
On the other hand, the $\phi-h-h$ coupling is more complicated than the SM
$h-h-h$ coupling. There is a momentum dependent part from the derivatives 
acting on the Higgs field, and this term can grow up as the radion mass gets 
larger or the CM energy gets larger in hadroproductions of the radion. It 
may lead to the violation of perturbative unitarity, which will be addressed
after we discuss the decay rates of the radion. Finally, the $h-\phi-\phi$ 
coupling could be described  by 
\begin{equation}
{\cal L}_{\rm int} (h \phi^2 ) = {v \over \Lambda_{\phi}^2} \phi^2~\left[  
{1 \over 2} \partial^2 h + {m_h^2 \over 2}~h \right],
\end{equation}
which might lead to an additional Higgs decay $h \rightarrow \phi\phi$,
if this mode is kinematically allowed, thereby enlarging the Higgs width 
compared to the SM case. However, this coupling actually vanishes upon using 
the equation of motion for the Higgs field $h$.
This is also in accord with the fact that the radion couples to the 
trace of the energy momentum tensor and there should be no $h-\phi$ mixing
after field redefinitions in terms of physical fields.  

In addition to the tree level $T_{\mu}^{\mu} ({\rm SM})^{\rm tree}$, 
there is also the trace anomaly term  for gauge fields \cite{trace} :
\begin{equation}
{T_\mu}^{\mu} ({\rm SM})^{\rm anom} = \sum_{G={\rm SU}(3)_C, \cdots } 
{\beta_G (g_G ) \over 2 g_G}~
{\rm tr} (F_{\mu\nu}^G F^{G \mu\nu}), 
\end{equation}
where $F_{\mu\nu}^G$ is the field strength tensor of the gauge group $G$ 
with generator(s) satisfying ${\rm tr}(t^a_G t^b_G)=\delta^{ab} $, and 
$\beta_G$ is the beta function for the corresponding gauge group. 
The trace anomaly term couples with the parameter of conformal transformation 
in our 3-brane. And the radion $\phi$ plays the same role as the parameter of 
conformal transformation, since it belongs to the warp factor in the 5 
dimensional RS metric \cite{csaki}. Therefore, the parameter associated with 
the conformal transformation is identified with the radion field $\phi$. As a 
result, the radion $\phi$ has a coupling to the trace anomaly term. For QCD 
sector as an example, one has
\begin{equation}
{\beta_{QCD} \over 2 g_s} = - (11 - {2 \over 3} n_f ) {\alpha_s \over 8 \pi}
\equiv - {\alpha_s \over 8 \pi}~b_{QCD},
\end{equation}
where $n_f = 6 $ is the number of active quark flavors. 
There are also counterparts in the $SU(2) \times U(1)$ sector.
This trace anomaly has an important phenomenological consequence.
For relatively light radion, the dominant decay mode will not be 
$\phi \rightarrow b \bar{b}$ as in the SM Higgs, but $\phi \rightarrow gg$.

Using the above interaction Lagrangian, it is straightforward to calculate 
the decay rates and branching ratios of the radion $\phi$ into 
$f \bar{f}, W^+ W^-, Z^0 Z^0, g g$ and $h h$. 
\begin{eqnarray}
\Gamma ( \phi \rightarrow f \bar{f} ) &=& N_c \frac{ m_f^2 m_{\phi}}
 { 8 \pi \Lambda_\phi^2} \left( 1 - x_f \right)^{3/2}, 
\nonumber 
\\
\Gamma( \phi \rightarrow W^+ W^- ) &=& \frac { m_\phi^3}
{16 \pi \Lambda_\phi^2} \sqrt{ 1 - x_W} \ ( 1 - x_W + \frac{3}{4} x_W^2 ), 
\nonumber   \\
 \Gamma( \phi \rightarrow Z Z ) &=& \frac { m_\phi^3}
{32 \pi \Lambda_\phi^2} \sqrt{ 1 - x_Z} \ ( 1 - x_Z + \frac{3}{4} x_Z^2 ),
\nonumber  \\
 \Gamma( \phi \rightarrow h h ) &=& \frac { m_\phi^3}{32 \pi 
\Lambda_\phi^2} \sqrt{ 1-x_h} \  ( 1 + \frac {x_h}{2} )^2,  
\nonumber  \\
\Gamma( \phi \rightarrow g g ) &=& \frac{ \alpha_s^2 m_\phi^3 }
{32 \pi^3 \Lambda_\phi^2} \left| b_{QCD} + \sum_q I_q( x_q ) \right|^2, 
\end{eqnarray}
where $ x_{f,W,Z,h} = 4 m_{f,W,Z,h}^2/m_\phi^2 $, and 
$I ( z ) = z [ 1 + (1 - z ) f(z) ]$ with
\begin{eqnarray} \label{ffnt}
f(z) & = &  - {1\over 2}~\int_0^1 {dy \over y}~\ln [ 1 - {4 \over z} y (1-y)]
\nonumber
\\
& = & \left\{  \begin{array}{cl}
           {\rm arcsin}^2(1/\sqrt{z}) \,,   &\qquad z \geq 1\,, \\
           -\frac{1}{4}\left[\ln \left(\frac{1+\sqrt{1-z}}{
                                             1-\sqrt{1-z}}\right)
                             -i\pi\right]^2\,, & \qquad z\leq 1\,.
               \end{array} 
\right.
\end{eqnarray}
Note that as $m_{t} \rightarrow \infty$, the loop function approaches
$I( x_t ) \rightarrow 2/3$ so that the top quark effect decouples and one
is left with $b_{QCD}$ with $n_f = 5$.  For $\phi \rightarrow WW,ZZ$, we have 
ignored $SU(2)_L \times U(1)_Y$ anomaly, since these couplings are allowed 
at the  tree level already, unlike the $\phi-g-g$ or $\phi-\gamma-\gamma$
couplings. This should be a good approximation for a relatively light radion.

\begin{figure}
\centerline{\epsfxsize=9cm \epsfbox{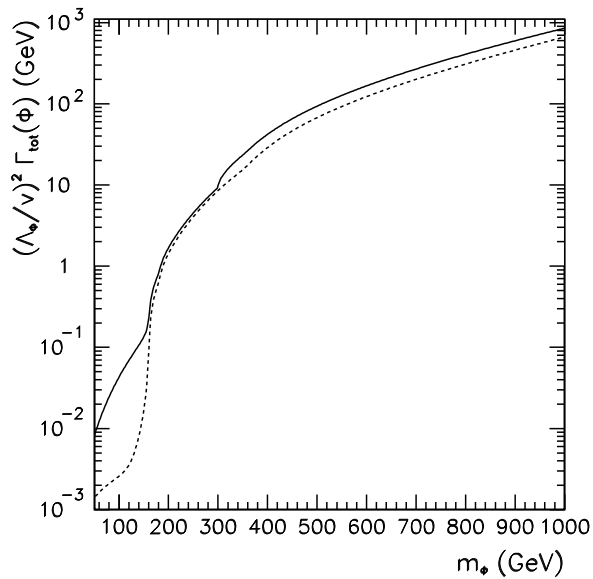}}  
\caption{The total decay rate (in GeV) for the radion $\phi$ for $m_h$=150 GeV
with a scale factor $(\Lambda_{\phi} / v )^2$.
The decay rate of the SM Higgs boson is shown in the dashed curve for 
comparison.
}
\label{decay}
\end{figure}

\begin{figure}
\centerline{\epsfxsize=9cm \epsfbox{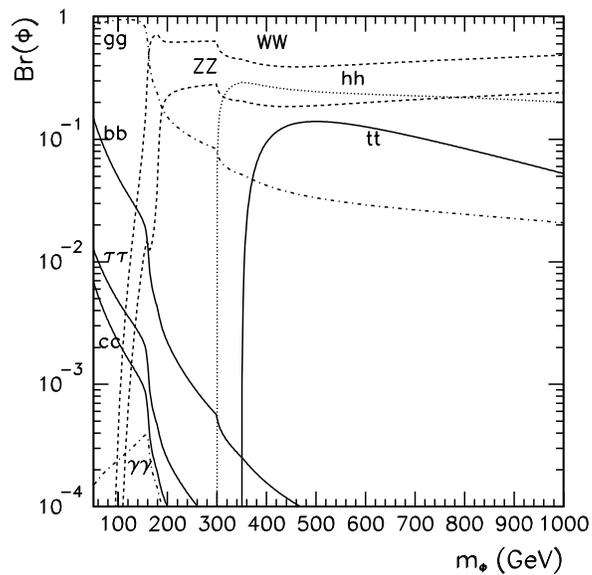}} 
\caption{The branching ratios for the radion $\phi$ into the SM particles.} 
\label{br}
\end{figure}

Using the above results, we show the decay rate of the radion and the 
relevant branching ratio for each channel available for a given $m_{\phi}$
in Figs.~\ref{decay} and \ref{br}.
In the numerical analysis, we use $\Lambda_{\phi} = v = 246$ GeV and 
$m_h = 150$ GeV, and also included the QCD corrections. The decay rates for 
different values of $\Lambda_{\phi}$ can
be obtained through the following  scaling : $\Gamma (\Lambda_{\phi})  
= (v/\Lambda_{\phi})^2 \Gamma (\Lambda_{\phi} = v)$. The decay rate scales 
as $(v/\Lambda_{\phi})^2$, but the branching ratios are independent of 
$\Lambda_{\phi}$.  In Fig.~\ref{decay}, we also show the decay rate of the 
SM Higgs boson with the same mass as $\phi$. We note that the light radion 
with $\Lambda_{\phi} = v$ could be a much broader resonance compared to the 
SM Higgs even if $m_{\phi} \lesssim 2 m_{W}$. This is because the 
dominant decay mode is $\phi \rightarrow g g $ (see Fig.~\ref{br}), unlike 
the SM Higgs for which the $b \bar{b}$ final state is a dominant decay mode. 
This phenomenon is purely a quantum field theoretical effect : enhanced 
$\phi-g-g$ coupling due to the trace anomaly. For a heavier radion, it turns 
out that $\phi \rightarrow V V $ with $V=W$ or $Z$ dominates other decay 
modes once it is kinematically allowed. Also the branching ratio for 
$\phi \rightarrow h h$ can be also appreciable if it is kinematically allowed.
This is one of the places where the difference between the SM and the radion 
comes in.
If $\Lambda_{\phi} \gg  v$, the radion would be a narrow resonance and should
be easily observed as a peak in the two jets or $WW(ZZ)$ final states.
Especially $\phi \rightarrow Z Z \rightarrow (l\bar{l}) (l^{'} \bar{l^{'}})$
will be a gold plated mode for detecting the radion as in the case of the SM
Higgs. Even in this channel, one can easily distinguish the radion from the 
SM Higgs by difference in their decay width. 

\begin{figure}
\centerline{\epsfxsize=9.5cm \epsfbox{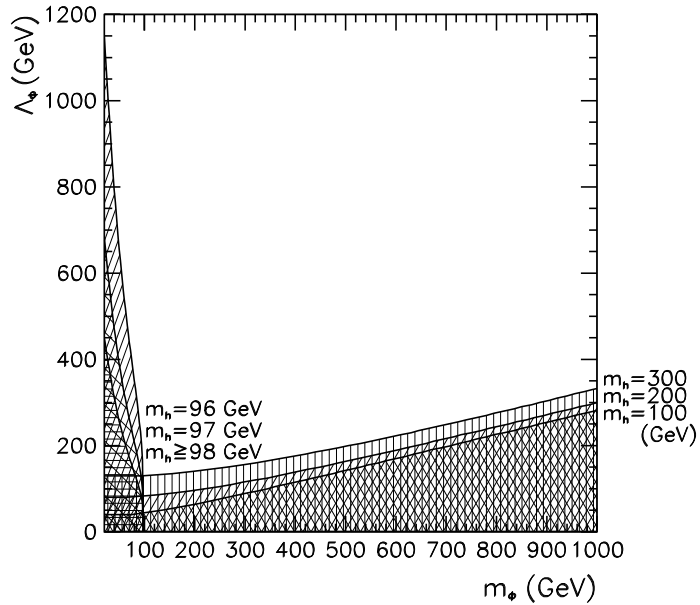}}
\caption{
The excluded region in the $m_\phi$ and $\Lambda_\phi$ space obtained from 
the recent L3 result on the SM Higgs search (the left three curves)  
and perturbative unitarity bound (the lower three curves). }
\label{bound}
\end{figure}

The perturbative unitarity can be violated (as in the SM) in the 
$V_L V_L \rightarrow V_L V_L$ or $h h \rightarrow V_L V_L$, etc. Here
we consider $h h \rightarrow h h$, since the $\phi-h-h$ coupling scales
like $s/\Lambda_{\phi}$ for large $s \equiv ( p_{h_1} + p_{h_2} )^2$.
The tree-level amplitude for this process is 
\begin{eqnarray}
{\cal M} (h h \rightarrow h h) &=& 
-\frac{1}{\Lambda_{\phi}^2} \left(
\frac{s^2}{s-m_{\phi}^2}+\frac{t^2}{t-m_{\phi}^2}+
\frac{u^2}{u-m_{\phi}^2} \right)
\nonumber
\\
&&-36\lambda^2 v^2 \left( \frac{1}{s-m_h^2}+ \frac{1}{t-m_h^2}
+\frac{1}{u-m_h^2} \right)
\nonumber
\\
&&-6\lambda,
\end{eqnarray}
where $\lambda$ is the Higgs quartic coupling, and $s+t+u = 4 m_h^2$.
Projecting out the $J=0$ partial wave component ($a_0$) and imposing
the partial wave unitarity condition $| a_0 |^2 \leq {\rm Im} (a_0)$ (i.e.
$| {\rm Re} (a_0)|\leq 1/2$), we get 
the following relation among $m_h, v, m_{\phi}$ and $\Lambda_{\phi}$, for 
$s \gg m_h^2,~ m_{\phi}^2$ :
\begin{equation}
\left|\frac{2m_h^2+m_{\phi}^2}{8\pi\Lambda_{\phi}^2}+\frac{3\lambda}{8\pi}
\right| \leq \frac{1}{2}.
\end{equation}
This bound is shown in the lower three curves of Fig.~\ref{bound}. We note 
that the perturbative unitarity is broken for relatively small 
$\Lambda_{\phi} \lesssim 130 (300)$ GeV for $m_{\phi} \sim$ 200 GeV (1 TeV).
Therefore, the tree level results should be taken with care for this range of
$\Lambda_{\phi}$ for a given radion mass. 

At the $e^+ e^-$ colliders, the main production mechanism for the radion 
$\phi$ is the same as the SM Higgs boson : the radion--strahlung from $Z$ 
and the $WW$ fusion, the latter of which becomes dominant for a larger CM 
energy \cite{ee}. Again we neglect the anomaly contributions here.
Since both of these processes are given by the rescaling of 
the SM Higgs production rates, we can use the current search limits  
on Higgs boson to get the bounds on radion.  
With the data from L3 collaboration\cite{l3}, we show the constraints of 
$\Lambda_\phi$ and $m_\phi$ in the left three curves of Fig.~\ref{bound}.
Since L3 data is for $\sqrt{s}=189$ GeV and mass of $Z$ boson is about 91 GeV, 
the allowed energy for a scalar particle is about 98 GeV. If the mass of the
scalar particle is larger than 98 GeV, then the cross section vanishes. 
Therefore if $m_\phi$ is larger than 98 GeV, there is no constraint on 
$\Lambda_\phi$. And the forbidden region in the $m_\phi-\Lambda_\phi$ plane
is not changed by $m_h \gtrsim 98$ GeV, because there is no Higgs 
contribution to the constraint for $m_h \gtrsim 98$ GeV. 

\begin{figure}
\centerline{\epsfxsize=9cm \epsfbox{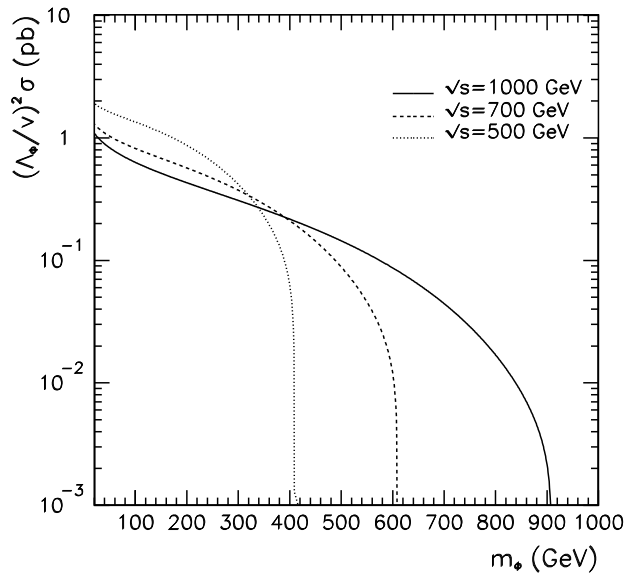}}
\caption{
The production cross section for the radion at NLC's at  
$\sqrt{s}$ = 500 , 700 and 1000 GeV, respectively.}
\label{lin}
\end{figure}

\begin{figure}
\centerline{\epsfxsize=9cm \epsfbox{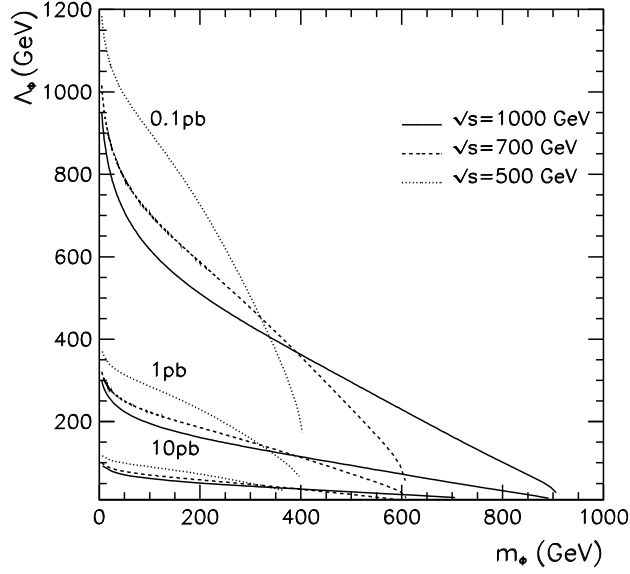}}
\caption{
The constant radion production cross section curves at next linear colliders
(NLC's)  for $\sqrt{s}$ = 500 , 700 and 1000 GeV }
\label{const-cs}
\end{figure}

The radion production cross sections at NLC's and the corresponding constant 
production cross section curves in the ($\Lambda_\phi$, $m_\phi$) plane are 
shown in Fig.~\ref{lin} and Fig.~\ref{const-cs}, respectively. We have chosen
three different CM energies for NLC's : $\sqrt{s} = 500$ GeV, 700 GeV and 1 
TeV. We observe that the relatively light radion ($m_{\phi} \lesssim 500$ GeV)
with $\Lambda_{\phi} \sim v$ (upto $\sim 1$ TeV) could be probed at NLC's 
if one can achieve high enough luminosity, since the production cross section 
in this region is less than a picobarn. 

\begin{figure}
\centerline{\epsfxsize=9cm \epsfbox{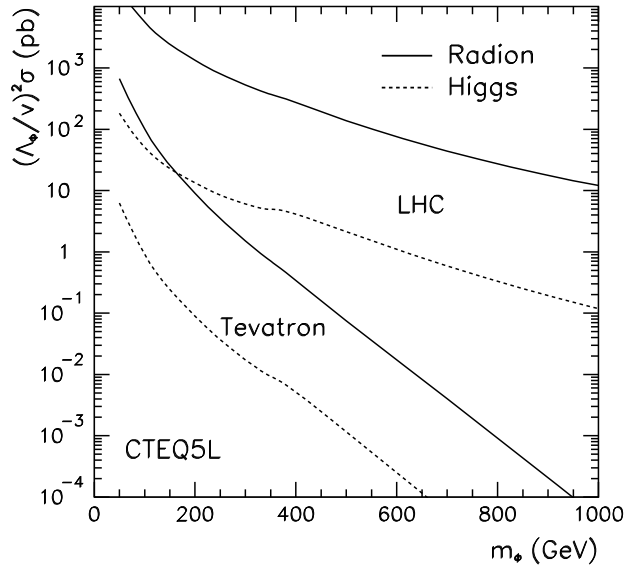}}
\caption{
The radion production cross section via gluon fusions at the Tevatron 
($\sqrt{s} = 2 $ TeV) and LHC ($\sqrt{s} = 14 $ TeV) with a scale factor 
$( \Lambda_{\phi} / v)^2$. The Higgs production cross sections are shown 
in dashed  curves for comparison. }
\label{hadron}
\end{figure}

The production cross sections of the radion at hadron colliders are given by
the gluon fusion into the radion through quark loop diagrams, as in the case
of Higgs boson production, and also through the trace anomaly term, Eq.~(4),
which is not present in the case of the SM Higgs boson :
\begin{equation}
\sigma ( p p \, ({\rm or}~ p \bar{p}) ) = K \, \hat{\sigma}_{\rm LO} 
( gg \rightarrow \phi ) \int_{\tau}^1 {\tau \over x}~g(x,Q^2) g(\tau /x,
Q^2) ~dx ,
\end{equation}
where $\tau \equiv m_{\phi}^2 / s$ and $\sqrt{s}$ is the CM energy of the 
hadron colliders ($\sqrt{s} = 2$ TeV and 14 TeV for the Tevatron and 
LHC, respectively). The $K$ factor includes the QCD corrections, and we set
$K=1.5$.
The parton level cross section for $gg\rightarrow \phi$ is given by 
\begin{equation}
\hat{\sigma}_{\rm LO} ( gg \rightarrow \phi ) = 
{\alpha_s^2 (Q) \over 256 \pi \Lambda_{\phi}^2} ~\left| b_{QCD} + 
\sum_q I_q ( x_q ) \right|^2, 
\end{equation}
where 
$I ( z )$ is given in the Eqs.~(\ref{ffnt}). For the gluon distribution 
function, we use the CTEQ5L parton distribution functions \cite{cteq5}.
In Fig.~\ref{hadron}, we show the radion production cross sections at the 
Tevatron and LHC as functions of $m_{\phi}$ for $\Lambda_{\phi} = v$. 
We set the renormalization scale $Q = m_{\phi}$ as shown in the figure. 
When we vary the scale $Q$ between $m_{\phi}/2$ and $2 m_{\phi}$, the 
production cross section changes about $+30 \%$ to $- 20 \%$. 
The production cross section will scale as $(v/\Lambda_{\phi})^2$ as before.
Compared to the SM Higgs boson productions, one can clearly observe that 
the trace anomaly can enhance the hadroproduction of a radion enormously. 
As in the SM Higgs boson, there is a great possibility to observe the radion 
upto mass $\sim 1 $ TeV if $\Lambda_{\phi} \sim v$.  For a smaller 
$\Lambda_{\phi}$, the  cross section becomes larger but the radion becomes 
very broader and it becomes more difficult to find such a scalar. For a larger
$\Lambda_{\phi}$, the situation becomes reversed : the smaller production
cross section, but a narrower width resonance, which is easier to detect.
In any case, however, one has to keep in mind that the perturbative unitarity
may be violated in the low $\Lambda_{\phi}$ region. 

In summary, we presented the collider phenomenology for the radion, which 
was suggested by means of stabilizing the modulus in Randall-Sundrum
scenario. Unlike other similar scenarios solving the hirerachy problem where
the radion or Kaluza-Klein modes are heavy and/or very weakly coupled to the 
SM fields, the radion discussed by Goldberger and Wise can have sizable 
interactions with the SM, only suppressed by a one power of the electroweak
scale. The radion phenomenology is very similar to the SM Higgs boson upto 
a simple rescaling of couplings by $v/\Lambda_{\phi}$, except that its 
couplings to two gluons or two photons are enhanced by the trace anomaly.
Also $\phi-h-h$ coupling can be substantially larger than the corresponding  
triple Higgs couplings, and it increases as $m_{\phi}$ or the CM energy 
grows up. We discussed various branching ratios and the decay rate of this 
radion, and the possibility to discover it at linear or hadron colliders.
Unlike the SM Higgs boson, the relatively  light radion dominantly decays 
into two gluons, not into the  $b \bar{b}$ pair, and this makes the radion
substantially broader than the SM Higgs if $\Lambda_{\phi} = v$. 
A heavier radion decays into $WW,ZZ$ pairs with some fraction into two Higgs 
if it is kinematically allowed. Depending on the $\Lambda_{\phi}$, the radion 
can be either broad or narrow, leading to larger (smaller) production rates
at hadron colliders but more harder (easier) to detectability. One may be 
also able to probe some regions of $(m_{\phi}, \Lambda_{\phi})$ if enough 
luminosity is achieved. 
It would be exciting to search for such a scalar particle which interacts 
with the trace of the energy-momentum tensor of the SM at the current/future
colliders. Finally, although we considered the radion in the 
Randall-Sundrum-Goldberger-Wise scenario, our study can be equally applied to 
any scalar particle which couples to the $T_{\mu}^{\mu} ({\rm SM})$.

\acknowledgements
The work of SB, PK and HSL was supported by grant No. 1999-2-111-002-5 from 
the interdisciplinary Research program of the KOSEF and BK21 project of the
Ministry of Education. JL is supported by the Alexander
von Humboldt Foundation.

\vspace{1.5cm}

{\it Notes Added} : 
While we were completing this work, there appeared two papers \cite{wells}
\cite{datta} that consider the same subjects as we do. The paper by Giudice 
{\it et al.} includes a direct coupling between the radion and the 
Ricci scalar parameterized by $\xi$ : ${\cal L}_{\rm int}= - \xi R \phi^2 /2$.
Our case corresponds to $\xi = 0$ in Ref.~\cite{wells}. 
And the usefulness of the mode $\phi\rightarrow \gamma\gamma$ at hadron
colliders was emphasized in Ref.~\cite{datta}.  
Qualitative conclusions of these papers are the same as ours where there are 
overlaps. 
 
\vfil\eject

\end{document}